\documentclass[12pt]{iopart}

\usepackage{graphicx}

\begin{document}

\title[]{AAMQS: a non-linear QCD description of new HERA data at small-x}

\author{ \underline{ Paloma Quiroga-Arias$^{1}$},  Javier L. Albacete$^{2}$, N\'estor Armesto$^{3}$, Jos\'e Guilherme Milhano$^{4,5}$, and Carlos A. Salgado$^{3}$}

\address{
$^1$ LPTHE, UPMC Univ. Paris 6 and CNRS UMR7589, Paris, France\\
$^2$ Institut de Physique Th\'eorique, CEA/Saclay, 91191 Gif-sur-Yvette cedex, France. URA 2306, unit\'e de recherche associ\'ee au CNRS.\\ 
$^3$ Departamento de F\'isica de Part\'iculas and IGFAE, Universidade de Santiago de Compostela 15706 Santiago de Compostela, Spain.\\
$^4$ CENTRA, Departamento de F\'isica, Instituto Superior T\'ecnico (IST),
Av. Rovisco Pais 1, P-1049-001 Lisboa, Portugal\\
$^5$ Physics Department, Theory Unit, CERN, CH-1211 Gen\`eve 23, Switzerland
}

\ead{pquiroga@lpthe.jussieu.fr}

\begin{abstract}
 We present a global analysis of available data on inclusive structure functions measured in electron-proton scattering at small values of Bjorken-x, including the latest data from the combined HERA analysis on reduced cross sections. Further, we discuss the kinematical domain where significant deviations from NLO-DGLAP should be expected and the ability of non-linnear physics to account for such deviations.
\end{abstract}


In the limit of small Bjorken-x, deviations from standard collinear perturbation theory are expected on account of large gluon densities. Therefore, non-linearities become relevant and need to be accounted for in the evolution. A solid theoretical framework in which to address the non-linear dynamics driving the small-$x$ evolution of the proton (nucleus) wave function is provided by the Colour Glass Condensate (CGC) \cite{Kovner:2000pt,Iancu:2000hn,Ferreiro:2001qy}, along with its B-JIMWLK evolution equations \cite{JalilianMarian:1997gr,Balitsky:1995ub}. On the phenomenological side, the implementation of saturation effects has proceeded via dipole-models in which the 
general physical properties of the CGC are implemented in an effective manner.

The Balitsky-Kovchegov equation~\cite{Balitsky:1995ub,Kovchegov:1999yj}, simplest theoretical realization of the CGC, fails to reproduce experimental data. However, it was demonstrated in~\cite{Albacete:2007yr} that considering only running coupling corrections accounts for most of the NLO effects. The ability of the rcBK equation to correctly describe experimental data was shown in \cite{Albacete:2009fh}. The rcBK equation, the most reliable and phenomenologically usable non-linear small-$x$ evolution tool, has become the standard choice for CGC computation of experimental observables (e.g., \cite{Albacete:2007sm,Albacete:2010bs,Albacete:2010pg}).

In the framework above described, a global fit of the inclusive DIS structure function led to the public release of a parametrization of the dipole scattering cross section at small values of $x$~\cite{Albacete:2009fh,Albacete:2010sy}, which we now proceed to present.

\section*{AAMQS setup}

The reduced DIS cross-section $\sigma_r$ can be written in terms of the virtual photon-proton cross-section $\sigma_{T,L}$, being  $y=Q^2/sx$ the inelasticity variable, as follows
\begin{equation}
\label{eq:sigred}
	 \sigma_r (x,y,Q^2) = \frac{Q^2}{4\,\pi^2\alpha_{em}}\Bigg(\sigma_{T} + \frac{2 (1-y)}{1+(1-y)^2} \sigma_{L}\Bigg)\, .
\end{equation}	 
 In the dipole formulation of QCD, valid for small-$x$, the photon-proton cross-sections for transverse and longitudinal polarization of the virtual photon can be expressed as
\begin{equation}
  \sigma_{T,L}(x,Q^2)=\sum_f \,  \sigma_{0,f}\, \int_0^1 dz \,d{\bf r}\,\vert
  \Psi_{T,L}^f(e_f,m_f,z,Q^2,{\bf r})\vert^2\,
  {\cal N}({\bf r},x)\,,
\label{dm1}
\end{equation}
where $\Psi_{T,L}^f$ is the light-cone wave function for a virtual photon to fluctuate into a $q\bar{q}$ dipole of quark flavor $f$ (with mass $m_f$ and electric charge $e_f$). ${\cal N}({\bf r},x)$ is the imaginary part of the dipole-target scattering amplitude averaged over impact parameter, with $\bf r$ the transverse dipole size. The transverse area over which quarks of a given flavour are distributed is represented by (twice) $\sigma_{0,f}$. Light quarks are taken to be identically distributed ($\sigma_{0,f=u,d,s} =\sigma_{0,light} $), while when accounting for heavy flavour contributions in (\ref{eq:sigred}) we allow $\sigma_{0,f=c,b} =\sigma_{0,heavy}$ to be different from $\sigma_{0,light}$. (see~\cite{Albacete:2010sy} for a more detailed explanation). The quantities $\sigma_{0,light}$ and $\sigma_{0,heavy} $ are free fit parameters. 

The evolution of  ${\cal N}({\bf r},x)$ is given by the rcBK equation:
\begin{eqnarray}
  \frac{\partial{\cal{N}}(r,x)}{\partial\,\ln(x_0/x)}&=& \int d{\bf r_1}\,
  K^{{\rm run}}({\bf r},{\bf r_1},{\bf r_2})\nonumber \\
 &\times&
 \left[{\cal N}(r_1,x)+{\cal N}(r_2,x)-{\cal N}(r,x)-
    {\cal N}(r_1,x)\,{\cal N}(r_2,x)\right]\, ,
\label{bkrun}
\end{eqnarray}
with $K^{{\rm run}}$ the evolution kernel including running coupling corrections~\cite{Balitsky:2006wa}. The running coupling in   $K^{{\rm run}}$  in (\ref{bkrun}) is evaluated at 1-loop accuracy in coordinate space
\begin{equation}
\label{eq:rc}
	\alpha_{s,n_f} (r^2) =  \frac{4\pi}{\beta_{0, n_f} \ln \Big(\frac{4 C^2}{r^2 \Lambda^2_{n_f}}\Big) }\, ,
	\qquad \beta_{0,n_f} = 11 - \frac{2}{3} n_f \, ,
\end{equation}
where the constant $C$ accounts for the uncertainty in the Fourier transform from momentum to coordinate space and is also a fit parameters. The coupling is evaluated with $n_f=3$ when only light quark contributions are taken into account, and with a variable flavour scheme once heavy flavours are included. Since in the rcBK equation  (\ref{bkrun}) all dipoles sizes are explored, an infrared regulation is called for: $\alpha_s (r^2 > r^2_{fr}) = \alpha_{fr}=0.7$ where $r^2_{fr}$ is the dipole size at which the coupling reaches $\alpha_{fr}$. 

We have explored in~\cite{Albacete:2010sy}  different families of initial conditions to solve~(\ref{bkrun}), finding that the results show a negligible dependence on the specific choice. For brevity, only results obtained with the GBW initial condition
\begin{equation}
	 {\cal N^{GBW}}(r,x\!=\!x_0)=
1-\exp{\left[-\frac{\left(r^2\,Q_{s\,0}^2\right)^{\gamma\,}}{4}\right] }\, ,
\end{equation}
are shown here. The fit parameters $Q_{s\,0}^2$ and $\gamma$ are, respectively, the saturation scale and the characteristic fall-off of the dipole scattering amplitude with decreasing $r$. 

\section*{Results}

\begin{figure}[h]
\begin{minipage}{7.5cm}
\includegraphics[width=1.0\linewidth,angle=0]{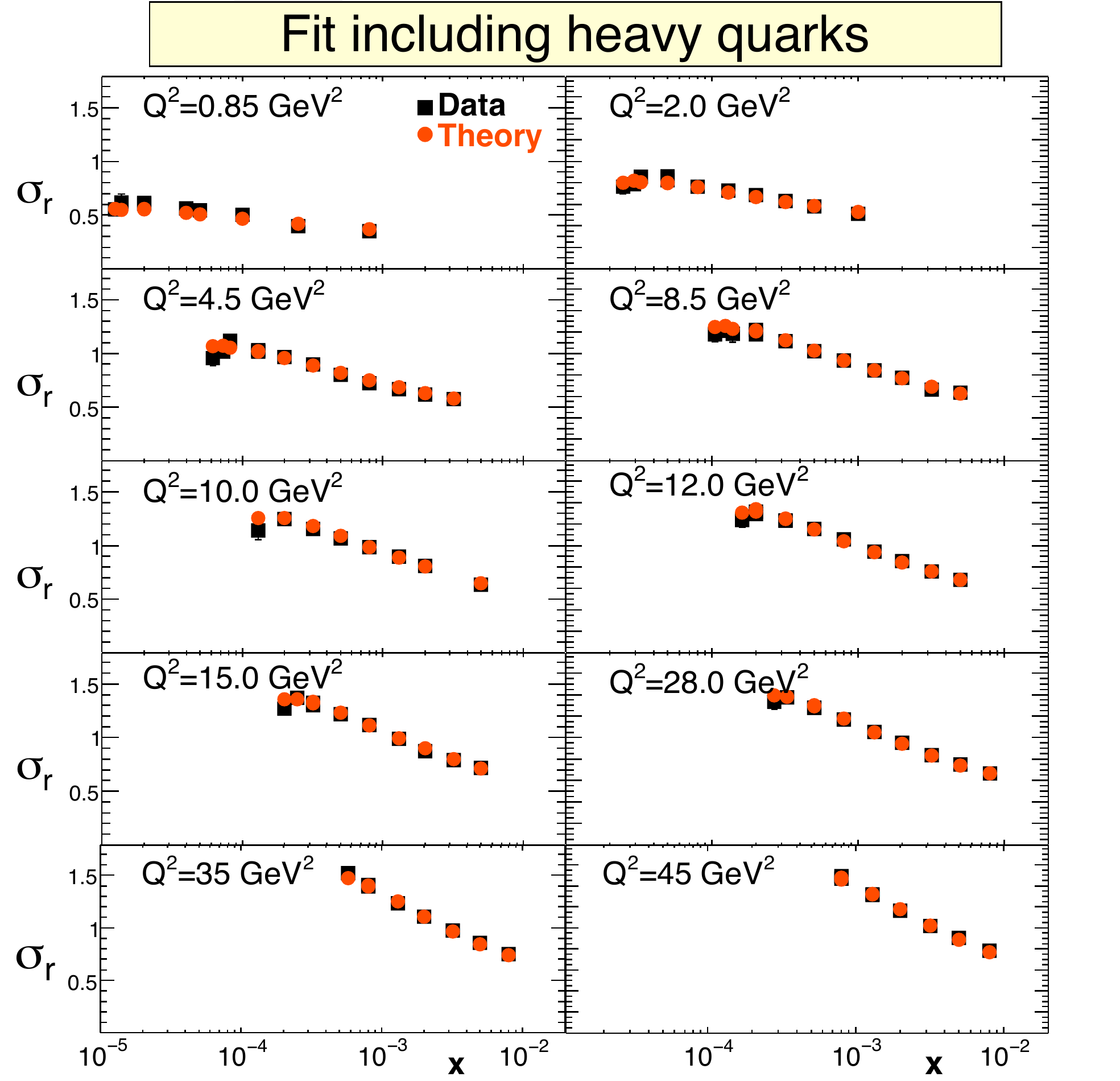}
\end{minipage}
\begin{minipage}{7.5cm}
\includegraphics[width=1.0\linewidth,angle=0]{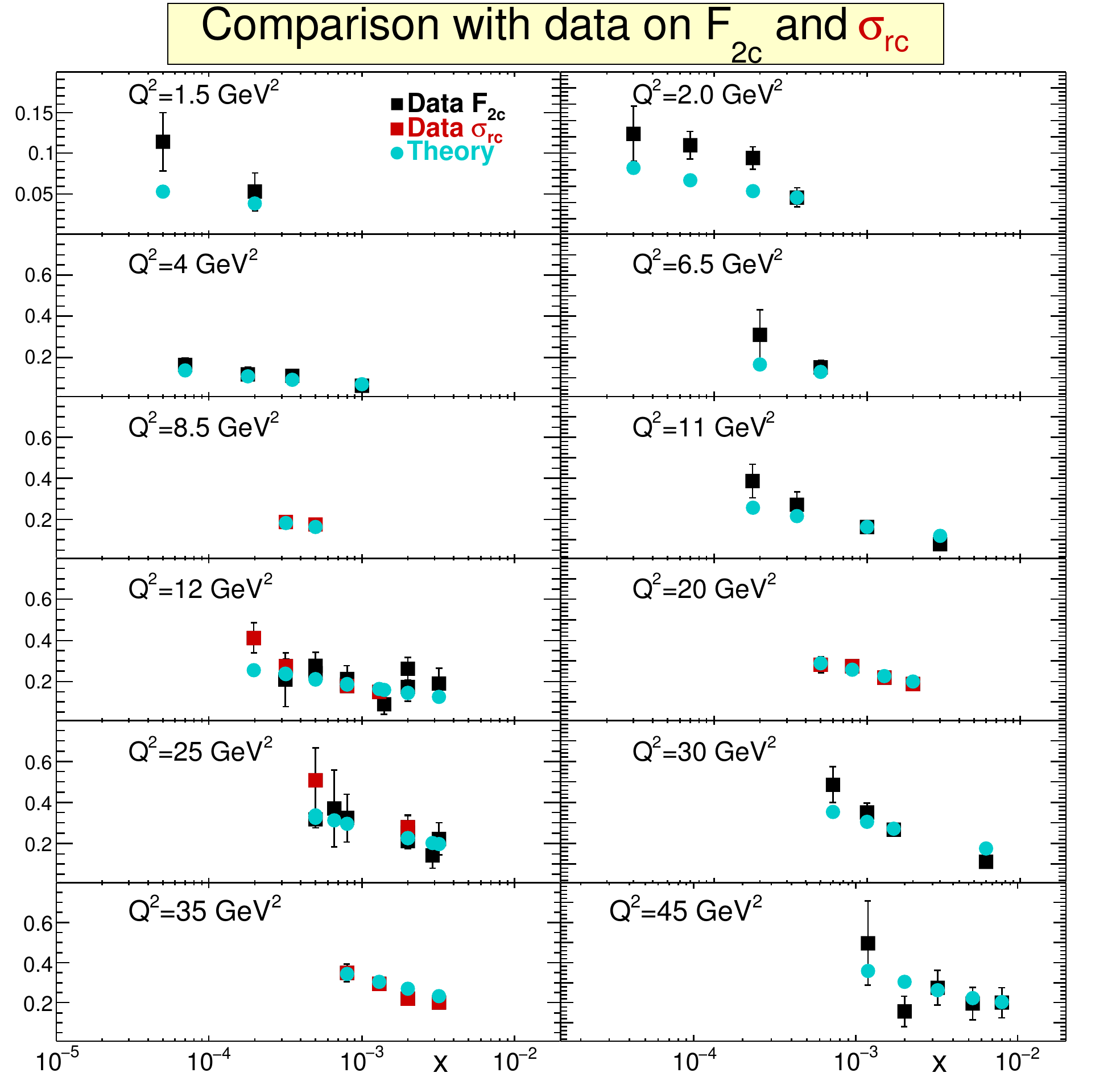}
\end{minipage}
\caption{\label{Fig1} (left) comparison of  $\sigma_r$ experimental data (black squares) with our fit results (red circles). (right) Coparison of experimental data for $F_{2c}$ (black squares), and $\sigma_{rc}$ (red squares) with our resuts (cyan circles) }
\end{figure}

In Fig.~\ref{Fig1} (left) we show the comparison of data for reduced cross section with the results of a fit to the combined H1/ZEUS data, E665 and NMC data for $F_2$, and available data on the charm contribution to $F_2$ and $\sigma_r$ within cuts $x<0.01$ and $Q^2<50$ GeV$^2$.
The parameters resulting from such fit were: $\chi^2/{\rm d.o.f.} =1.30$,  $Q_{s\,0, light(heavy)}^2 =0.23(0.22)$GeV$^2$, $\sigma_{0(c)}=36.36(20.38)$mb, $\gamma_{(c)} =1.24(0.92)$, C=7.86.  Notice the remarkable agreement of the AAMQS calculation with the experimental data despite their high accuracy (errors are often smaller than the size of the symbols used for the data points).
In Fig.~\ref{Fig1} (right) we present a comparison to the charm component of $F_{2,c}$ and $\sigma_{r,c}$ with the results of the fit described above, showing a good description of the experimental data as well.

\begin{figure}[h]
\begin{minipage}{9cm}
\includegraphics[width=1.1\linewidth,angle=0]{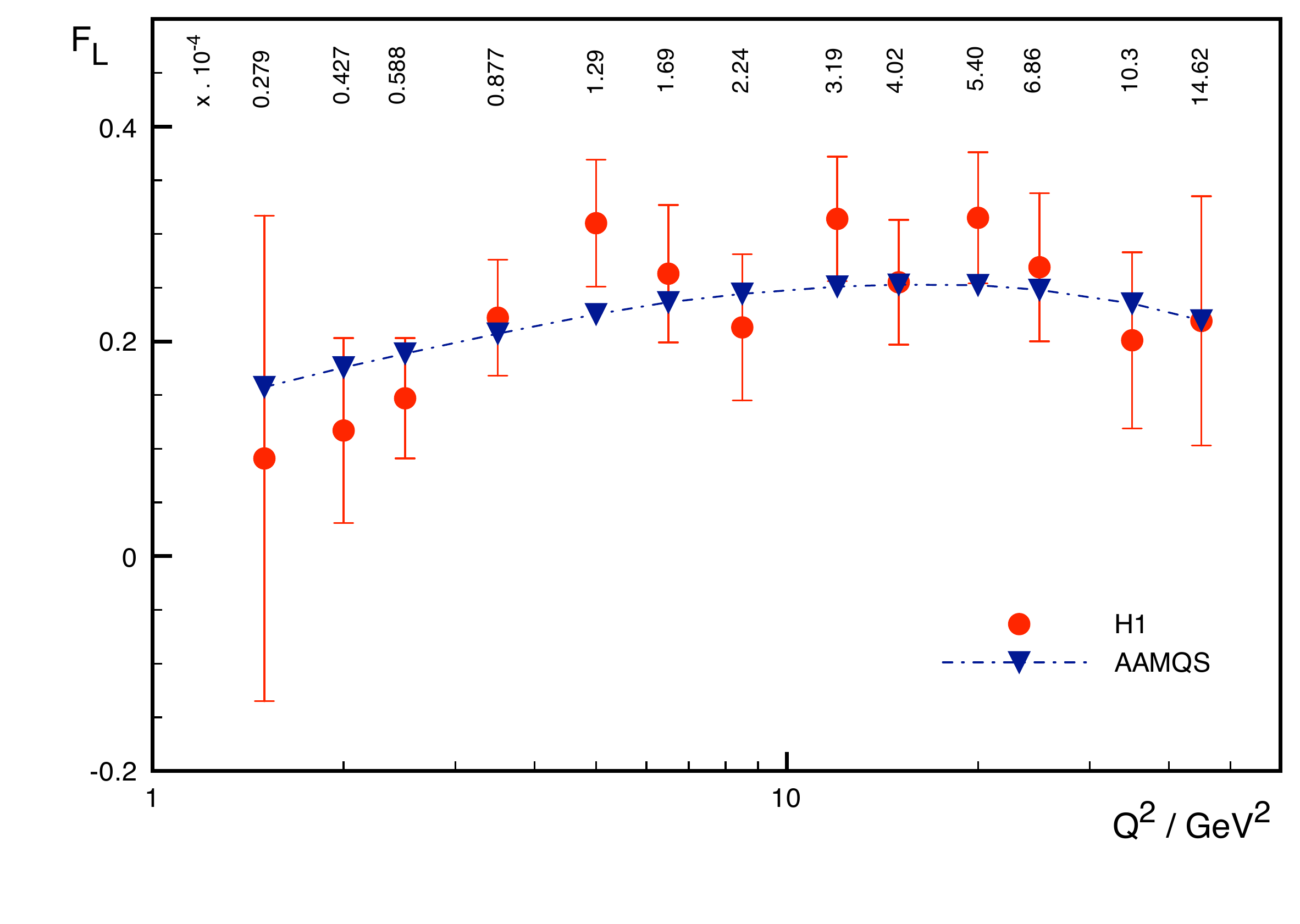}
\end{minipage}
\hspace{-1.2cm}
\begin{minipage}{8cm}
\caption{\label{Fig2} Comparison of  the latest $F_{L}$ data~\cite{Collaboration:2010ry} (red dots)  with the AAMQS calculation using the results of the fit shown if Fig.~\ref{Fig1} (blue triangles). }
\end{minipage}
\vspace{-2pc}
\end{figure}

As an independent check, since no data of this sort are included in any of the fits, we calculate the longitudinal structure function $F_L$ and compare our result to the latest H1 measurement~\cite{Collaboration:2010ry}. The result is shown in Fig.~\ref{Fig2} as a function of $Q^2$ for the different values of $x$ finding an excellent agreement.

Finally, we discuss the kinematical domain where significant deviations from NLO-DGLAP should be expected. In~\cite{Caola:2009iy} it was shown that there is tension in NLO-DGLAP fits to HERA data on structure functions when data sets with $Q^2$ below the estimated saturation scale of the proton were excluded from the analysis. Within the AAMQS framework a similar exercise can be performed by excluding from the fits all data above a given $x=x_{cut}$, and then extrapolating to the unfitted region.  Since linear and non-linear effects are expected to be present in  different (and complementary) regions of phase space, we need to determine where the saturation boundary is. To do so, in~\cite{aamqs-nnpdf} both linear and non-linear fits to partial sets of data are performed in their respective safe ($x,Q^2$) regions, leaving a common unfitted region in which the extrapolation from both formalisms is compared. We find that the deviations from NLO-DGLAP reported in~\cite{Caola:2009iy} do not appear when the non-linear rcBK fit is performed at low-x and extrapolated to the unfitted region. This suggests  the presence of saturation effects which are not accounted for in linear evolution.

\bigskip
\noindent
\textbf{ACKNOWLEDGMENTS} We acknowledge support from MICINN (Spain) under project FPA2008-01177; Xunta de Galicia (Conselleria de Educacion) and through grant PGIDIT07PXIB206126PR; the Spanish Consolider- Ingenio 2010 Programme CPAN (CSD2007-00042); Funda\c c\~ao para a Ci\^encia e a Tecnologia (Portugal) under project CERN/FP/109356/2009 (JGM); and the French ANR under contract ANR-09-BLAN-0060 (PQA).

\section*{References}
\bibliography{bib}

\end{document}